\documentclass[12pt,preprint]{aastex}

\shorttitle{Prediction of Solar Cycles}
 \shortauthors{Kitiashvili and Kosovichev}

\begin{document}

\title{Application of Data Assimilation Method for Predicting Solar Cycles}

\author{I. Kitiashvili}
\affil{Center for Turbulence Research, Stanford University, Stanford, CA 94305, USA}
\email{irinasun@stanford.edu}

\and

\author{A. G. Kosovichev}
\affil{Hansen Experimental Physics Laboratory, Stanford
University, Stanford, CA 94305, USA}
\email{sasha@sun.stanford.edu}

\begin{abstract}
Despite the known general properties of the solar cycles, a reliable
forecast of the 11-year sunspot number variations is still a
problem. The difficulties are caused by the apparent chaotic
behavior of the sunspot numbers from cycle to cycle and by the
influence of various turbulent dynamo processes, which are far from
understanding. For predicting the solar cycle properties we make an
initial attempt to use the Ensemble Kalman Filter (EnKF), a data
assimilation method, which takes into account uncertainties of a
dynamo model and measurements, and allows to estimate future
observational data. We present the results of forecasting of the solar
cycles obtained by the EnKF method in application to a low-mode
nonlinear dynamical system modeling the solar $\alpha\Omega$-dynamo
process with variable magnetic helicity. Calculations of the
predictions for the previous sunspot cycles show a reasonable agreement
with the actual data. This forecast model predicts that the next sunspot
cycle will be significantly weaker (by $\sim 30\%$) than the previous cycle,
continuing the trend of low solar activity.
\end{abstract}


\section{Introduction}
Investigation of solar activity is one of the oldest solar physics
problems. Here we consider this phenomenon in the context of the
sunspot number variations, which have detailed observational data
during the past 23 solar cycles. There is no doubt that the 11-year
cyclic variations of the sunspot number are connected to the dynamo
process inside the Sun. Therefore it is natural to search for a
solution by investigating nonlinear dynamo models, and use them for
predicting the solar cycles. However, our current understanding of
the solar dynamo is quite poor. The current predictions of the next
solar cycle, number 24, show a wide range of the expected sunspot
number. For example, using an axisymmetric, mean-field model of a
flux transport dynamo Dikpati and Gilman (2006) predicted that the
cycle 24 will be about 30\% - 50\% stronger than the previous cycle
23 with the sunspot number reaching 155 - 180 at the maximum.
Using polar magnetic field measurements and assuming a direct relationship
between the polar field strength and the future toroidal magnetic field Svalgaard et al. (2005)
predicted that the cycle 24 will have a peak sunspot number of $75 \pm 8$.
A weak cycle 24 with the sunspot number of $74 \pm 10$ was also proposed
by Javaraiah (2007) from analysis of sums of the areas of sunspot groups
in the  latitudinal interval of $0^\circ - 10^\circ$ in both hemispheres.
There are many other predictions for the upcoming cycle(s), reviewed by Kane (2007)
and Obridko and Shelting (2008).

The great variety of the predictions is caused by uncertainties in
the dynamo models and model parameters, and errors in both models
and observations. In the paper, we present initial results of
application of a data assimilation method to a simple nonlinear dynamo model
\citep{kitikos08}. This model reproduces the basic properties of a
solar cycle, such as the shape of the sunspot number profile. The
advantage of the data assimilation methods is in their ability to
combine the observational data and the models for possible efficient
and accurate estimations of the physical properties, which cannot be
observed directly. Here we consider an implementation of the
Ensemble Kalman Filter method, EnKF, \citep{evensen07}, which is
effective for investigation of nonlinear dynamical models. The
method is useful in several aspects: it supports estimations of past,
present, and even future states of a system, and it can do so even when
the precise nature of a modeled system is unknown.

\section{Basic formulation of the data assimilation method}

The main goal of any model is an accurate description of properties
of a system in the past and present times, and the prediction of its
future behavior. However, a model is usually constructed with some
approximations and assumptions, and has errors. Therefore, it cannot
describe the true condition of a system. On the other hand,
observational data, $d$, also include errors, $\epsilon$, which are
often difficult to estimate. The data assimilation methods such as
the Kalman Filter \citep{kalman60} allow us, with the help of an
already constructed model and observational data, to determine the
initial state of the model, which will be in agreement with a set of
observations, and obtain a forecast of future observations and
estimate its errors \citep{evensen07, kiti08}. For instance, in
our case we know from observations the sunspot number (with some
errors) and want to estimate the state of the solar magnetic fields,
described by a dynamo model.

In generally, if the state, $\psi$, of a system can be described by a dynamical model
$d\psi/dt = g(\psi,t)+q$, with initial conditions $\psi_{0} =
\Psi_{0}+p$, where $g(\psi,t)$ is a nonlinear vector-function, $q$
and $p$ are the errors of the model and in the initial conditions,
then the system forecast is $\psi^{f}=\psi^{t}+ \phi$, where
$\psi^{t}$ is the true system state, and $\phi$ is the forecast error.
The relationship between the true state and the observational data is
given by a relation $d= M[\psi]+ \epsilon$, where $d$ is a vector of
measurements, $M[\psi]$ is a measurement functional, which relates
the model state, $\psi$, to the observations, $d$.

For a realization of the data assimilation procedure in the case of
nonlinear dynamics it is convenient to use the Ensemble Kalman
Filter (EnKF) method \citep{evensen07}. The main difference
of the EnKF from the standard Kalman Filter is in using for the
analysis an ensemble of possible states of a system, which can be
generated by Monte Carlo simulations. If we have an ensemble of
measurements $d_{j} = d + \epsilon_{j}$ with errors $\epsilon_{j}$
(where $j = 1, ..., N$) then we can define the covariance matrix
of the measurement errors $C_{\epsilon\epsilon}^{e} =
\overline{\epsilon\epsilon^{T}}$, where the over bar means the
ensemble averaged value, and superscript $T$ indicates
transposition. Using a model we always can describe future states of
a system, $\psi^{f}$. However, errors in the model, initial
conditions and measurements do not allow the model result be consistent
with observations. To take into account this deviation, we consider a
covariance matrix of the first guess estimates (our forecast
related only to model calculations): $(C_{\psi\psi}^{e})^{f} =
\overline{(\psi^{f}-\overline{\psi^{f}})(\psi^{f}-\overline{\psi^{f}})^{T}}$.
Note, that the covariance error matrix is calculated for every
ensemble element. Then, the estimate of the system state is given
by:
\begin{eqnarray}
\psi^{a} = \psi^{f}+K \left(d - M \psi^{f}\right),
\end{eqnarray}
where $K =
(C_{\psi\psi}^{e})^{f}M^{T}\left(M(C_{\psi\psi}^{e})^{f}M^{T}
+C_{\epsilon\epsilon}^{e}\right)^{-1}$, is the so-called Kalman gain
\citep{kalman60,evensen07}. The covariance error matrix of the best
estimate is calculated as: $(C_{\psi\psi}^{e})^{a}
=\overline{(\psi^{a}-\overline{\psi^{a}})(\psi^{a}-
\overline{\psi^{a}})^{T}}=
\left(I-K_{e}M\right)(C_{\psi\psi}^{e})^{f}$. We can use the last
best estimate obtained with the available observational data as the
initial conditions and make the next forecast step. At the forecast
step, we calculate a reference solution of the model, according to
the new initial conditions, then simulate measurements by adding
errors to the model and to the initial conditions. Finally we obtain
a new best estimate of the system state, which is our forecast.
A new set of observations allows us to redefine the
previous model state and make a correction to the predicted state.

In order to implement EnKF for prediction of the sunspot cycles it
is necessary to define a dynamo model, which describes the evolution
of the system parameters in time.

\section{Dynamo model}

Currently, there is no generally accepted model of the solar dynamo.
However, most of the models are based on the Parker's oscillatory
$\alpha\Omega$-dynamo mechanism \citep{parker55}, which includes
turbulent helicity and magnetic field stretching by the differential
rotation. Recent observational and theoretical investigations
\citep[e.g.][]{sokol07,brand05} revealed an important role of
magnetic helicity \citep{pouquet76}. Thus, for this investigation we
added to the original Parker's model an equation describing the
evolution of the magnetic helicity, $\alpha_{m}$. This equation was
derived by Kleeorin and Ruzmaikin (1982) from the conservation of
the total magnetic helicity. Then, the dynamo model can be written
as \citep{kitikos08}
\begin{eqnarray}
\frac{\partial A}{\partial t}&=&\alpha B+\eta \nabla ^{2}A, \,\,
\frac{\partial B}{\partial t}=G\frac{\partial A}{\partial x}+\eta
\nabla ^{2}B,\\ \nonumber
 \frac{\partial \alpha_{m}}{\partial t}&=&\frac{Q}{2\pi
\rho} \left[\langle{\vec B}\rangle(\nabla \times \langle{\vec
B}\rangle)-\frac{\alpha}{\eta} \langle{\vec
B}\rangle^{2}\right]-\frac{\alpha_{m}}{T},
\end{eqnarray}
where $B$ is the toroidal component of magnetic field, $A$ is the
vector potential of the poloidal component of the mean magnetic
field, $\langle\vec B\rangle=\vec B_{P}+\vec B_{T}$ ($\vec
B_{P}={\rm curl}(0,0,A)$, $\vec B_{T}=(0,0,B)$ in spherical
coordinates), $\eta$ describes the total magnetic diffusivity, which
is the sum of the turbulent and molecular magnetic diffusivity,
$\eta = \eta_{t}+\eta_{m}$ (usually $\eta_{m} << \eta_{t}$); $G =
\partial\left<v_x\right>/\partial y$ is the
rotational shear, coordinates $x$ and $y$ are in the azimuthal and
latitudinal directions respectively, parameter $\alpha$ is helicity
represented in the form $\alpha=\alpha_{h}/(1+\xi
B^{2})+\alpha_{m}$, $\alpha_{h}$ and $\alpha_{m}$ are the kinetic
and magnetic parts; $\xi$ is a quenching parameter, $\rho$ is
density, $T$ is a characteristic time of dissipation magnetic
helicity (which includes dissipation though helicity transport) and,
$Q \sim 0.1$.

Following the approach of Weiss {\it et al.} (1984) we average the
system of equations (2) in a vertical layer to eliminate
$z$-dependence of $A$ and $B$ and consider a single Fourier mode
propagating in the $x$-direction assuming $A=A(t)e^{ikx}$,
$B=B(t)e^{ikx}$; then we get the following system of equations
\begin{eqnarray}
\frac{{\rm d} A}{{\rm d} t}&=&\alpha B-\eta k^{2} A, \qquad
\frac{{\rm d} B}{{\rm d} t}=ikG A-\eta k^{2}B,\\
\frac{{\rm d} \alpha_{m}}{{\rm d} t}&=&-\frac{\alpha_{m}}{T}-\frac{Q}{2\pi \rho}
\left[-ABk^{2}+\frac{\alpha}{\eta}\left(B^{2}-k^{2}A^{2}\right)\right]. \nonumber
\end{eqnarray}

For the interpretation of the solutions of the dynamical system in
terms of the sunspot number properties we use the imaginary part of
the toroidal component $B(t)$ because it gives the amplitude of the
antisymmetric harmonics, and approximate the sunspot number, $W$, as
$({\rm Im}B)^{3/2}$, following Bracewell's suggestion
\citep{brace53, brace88}. This dynamo model has been investigated in
detail by Kitiashvili \& Kosovichev (2008).

\begin{figure}
\begin{center}
\includegraphics[scale=0.7]{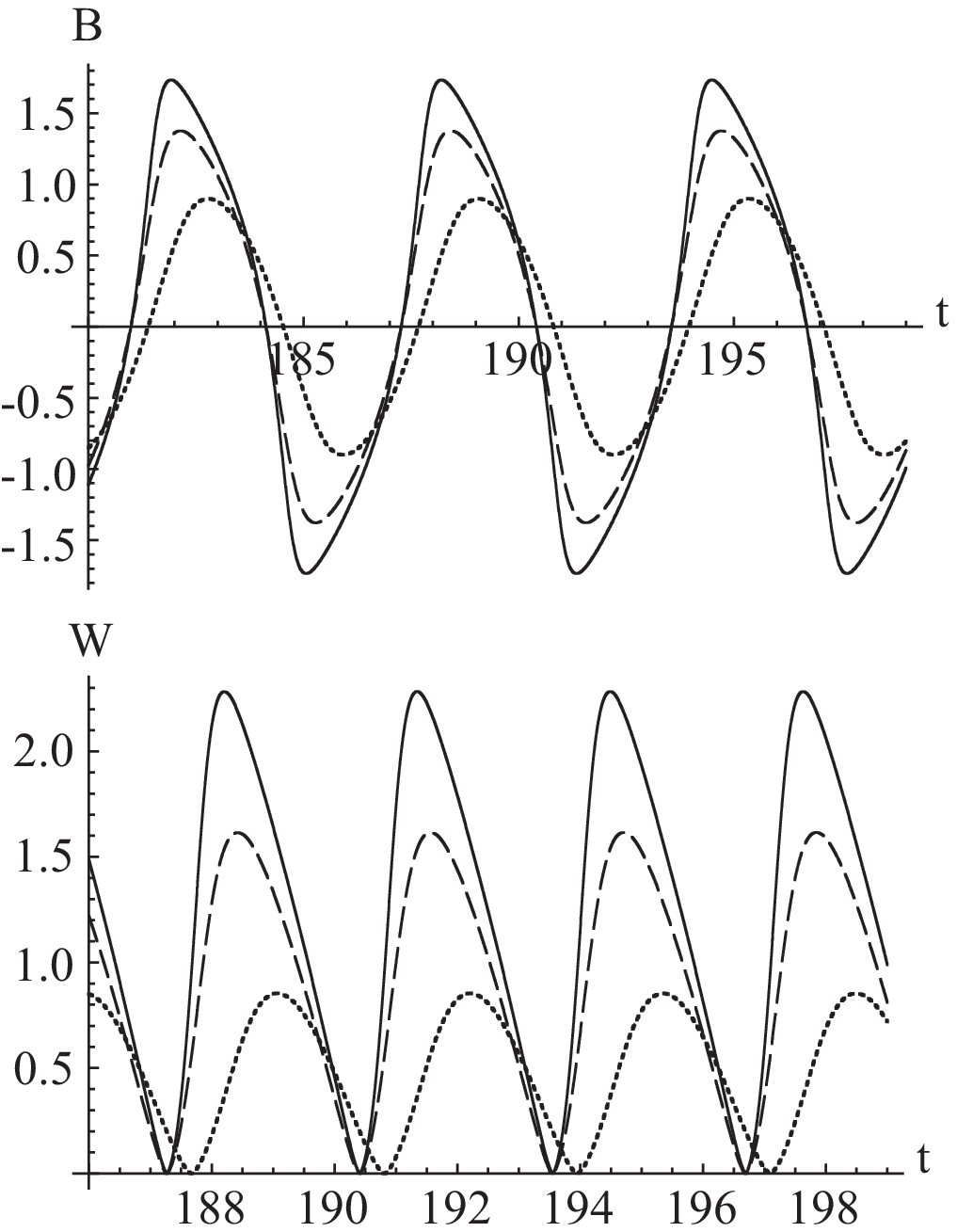}
\end{center}
\caption{A typical periodic solution (in non-dimensional units) of
the dynamo model:  toroidal magnetic field, $B$ (top panel), and the
model sunspot number, $W$ (bottom panel)
\citep{kitikos08}. \label{fig1}}
\end{figure}

Figure \ref{fig1} shows typical nonlinear periodic solutions and the
corresponding model sunspot number, which reproduce typical observed
solar cycle profiles with fast growth and  slow decay. The profile
of the toroidal field variations becomes close to the sinusoidal
behavior for small amplitude. Perhaps, it is essential that the
model gives a qualitatively correct relationship between the model
sunspot number amplitude and the growth time.

\section{Implementation of Ensemble Kalman Filter}

For the assimilation of the sunspot data into the dynamo model we
selected a class of periodic solutions, which corresponds to
parameters of the  middle convective zone and describes the typical
behavior of the sunspot number variations (Fig.~\ref{fig1}). The
implementation of the EnKF method consists of 3 steps: preparation
of the observational data for analysis, correction of the model
solution according to observations, and prediction.

{\it Step 1: Preparation of the observational data.} Following
Bracewell (1953, 1988), we transform the annual smoothed values of
the sunspot number for the period of 1856 - 2007 in the toroidal
field values using the relationship $B\sim W^{2/3}$ and alternating
the sign of $B$. Also we select the initial conditions of the model
so that the reference solution coincides with the beginning of the
first cycle of our series, cycle 10, which started in 1856.
In this paper we do not consider the previous solar cycles because
of the uncertainties in the early sunspot number measurements \citep{svalg07}.
Then we normalize the toroidal field in the model in
such a way that the model amplitude of $B$ is equal to the mean
''observed" toroidal field. In addition, we normalize the model time
scale assuming that the period of the model sunspot variations
corresponds to the typical solar cycle duration of 11 years.

\begin{figure}
\begin{center}
\includegraphics[scale=1.1]{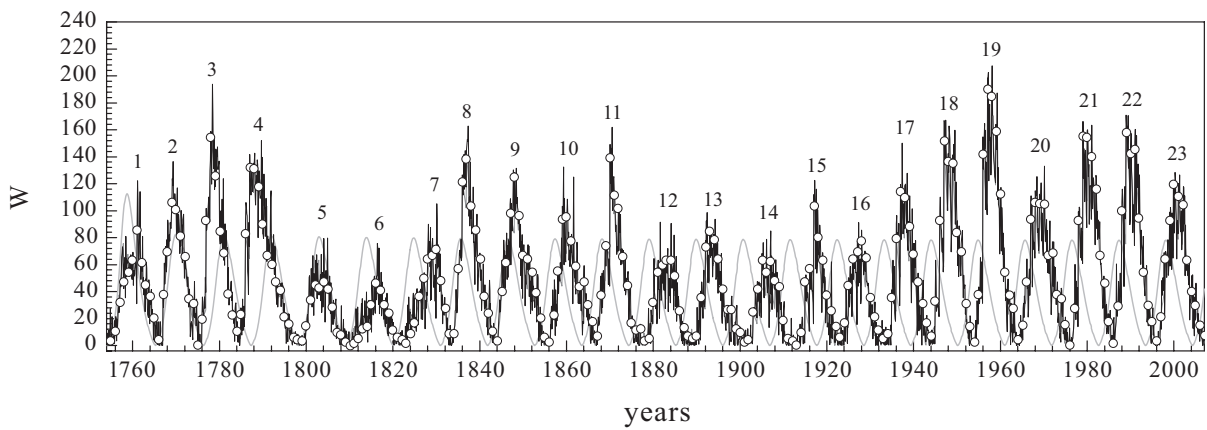}
\end{center}
\caption{Results of assimilation of the annual sunspot number data
(circles) into the dynamo model. The grey curve shows the reference
solution (without assimilation analysis), and the black curve shows
the best EnKF estimate of the sunspot number variations, obtained
from the data and the dynamo model. \label{fig2}}
\end{figure}

{\it Step 2: Assimilation for the past system state.} Unfortunately
we do not have observations of the magnetic helicity, toroidal and
poloidal components of magnetic field. Therefore, in the first
approximation, we generate observational data as random values
around the reference solution with a standard deviation of $\sim
12\%$, which was chosen to roughly reproduce the observed variations of the
sunspot number. Similar random errors are also added to the model
equations as described in Sec.~2. Then, we calculate the covariance
error matrixes of the observations, $C_{\epsilon\epsilon}^{e}$, and
the forecast, $(C_{\psi\psi}^{e})^{f}$.  After combining the
observational and model error covariances in the form of Kalman
gain, $K$, we obtain the best estimate of evolution of the system,
$\psi^{a}$ from Eq.~1. Figure \ref{fig2} shows the result of
assimilation of the sunspot data into the dynamo model: the best
EnKF estimate (black curve), the initial model (grey curve) and the
actual sunspot data (circles).

{\it Step 3: Prediction.} For obtaining a prediction of the next
solar cycle we determine the initial conditions from the best
estimated solution for the previous cycle in terms of the amplitude and phase
to continue the model calculations. Then after receiving the reference
solution with the new initial conditions we simulate future observational
data by adding random noise and repeat the analysis. This provides the best
EnKF estimate of the future state of the system (forecast).

The described analysis has been tested by calculating predictions of
the previous cycles. Figure \ref{fig3} (a-h) shows the examples of the
EnKF method implementation for forecasting the sunspot number of cycles 16-23.
For these forecasts, we first obtain the best estimated solutions (blue curves)
using the observational data prior to these cycles (open circles). After this,
we obtain the exact solution (black curves) according to the initial conditions
of the time of the last measurement and simulate a new set observation (black dots)
by adding random noise. Then, we obtain the EnKF estimates using the simulated
observations, which give us the prediction (Fig.~\ref{fig3}, red curves).
These experiments show that this approach can provide reasonable forecast of the
strength of the next solar cycles. However, there are significant discrepancies.
For instance, the strength of the cycle 16 is overestimated, and the strength
of cycle 19 is underestimated. The main uncertainties are caused by in accuracies
in determining the time of the end of the previous cycle from the sunspot number
data, and, of course, by the incompleteness of the model and insufficiency
the sunspot number data. In particular, we found the forecast is inaccurate
when the sunspot number change significantly from the value of the previous cycle.
Also, our forecast experiments showed a strong dependence on the phase relation between
the reference model solution and the observations. The phase difference appears due
to the constant period of the solution the variations in the duration of the
solar cycles. Curiously, when the model phase is ahead of the solar cycle phase then
adding a data point at the start of a cycle substantially improves the forecast.
However, when the model phase lags thai does not happen. This effect is taken
into account by correcting the phase of the reference solution that it is slightly
ahead of the solar cycle phase.

The same analysis scheme is applied for prediction of the next solar
cycle 24 (Fig.~\ref{fig3}i). According to this result, the solar
cycle 24 will be weaker than the current cycle by approximately
30\%. To check the stability of the prediction we used two other
sets of initial conditions in 2008 and obtained close results (Fig.~\ref{fig3}i).

\section{Discussion and Conclusion}

\begin{figure}
\includegraphics[scale=1.15]{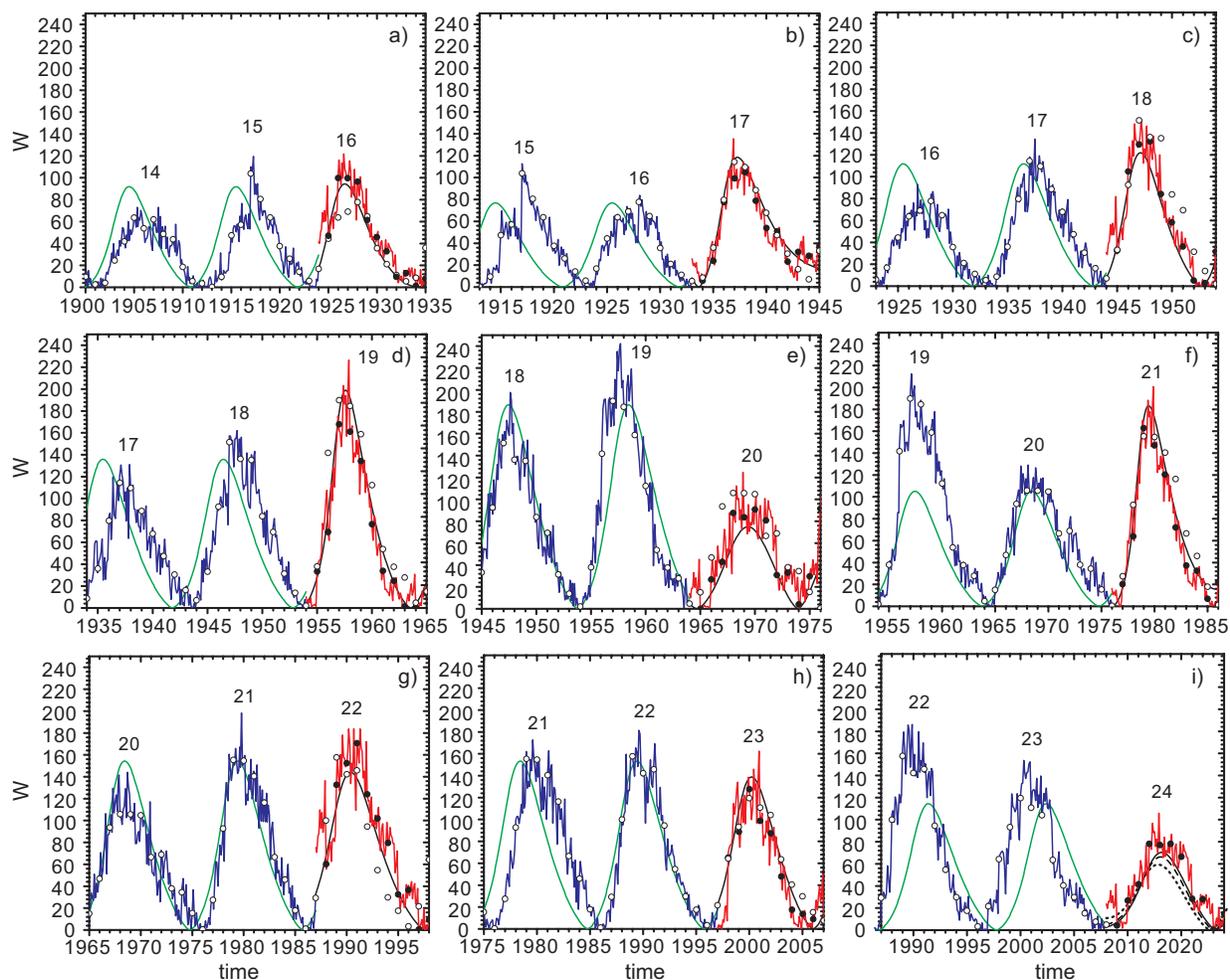}
\caption{Predictions for the solar cycles 16-24. The green curves show
the model reference solution. The blue curves show the best estimate of the
sunspot number using the observational data (empty circles) and the model,
for the previous cycles. The black curve the model solution according to the
initial conditions of the last measurement. The red curves show the prediction
results. In panel i) the model solution is shown for 3 different estimates of
the sunspot number for 2008: 3 (black curve), 5 (dashes) and 10 (dots).\label{fig3}}
\end{figure}

The results of assimilation of the annual sunspot number data into
the solar dynamo model and the prediction of the previous
solar cycles (Fig.~\ref{fig3}) demonstrate a new method of
forecasting the solar activity cycles. Using the EnKF method and a
simple dynamo model we obtained reasonable predictions usually for the
first halfs of the sunspot cycles with the error $\sim 8 - 12$\%,
and in some cases also for the declining phase of the cycles.
This method predicts a weak solar cycle 24 (Fig.~\ref{fig3}i)
with the smoothed annual sunspot number at the maximum of approximately
80. It is interesting that the simulations show that the previous cycle
does not finish in 2007 as was expected, but it still continues.
According to the estimates the maximum of the next cycle will be
approximately in 2013.

The application of the data assimilation method, EnKF, for modeling and
predicting the solar cycles shows the power of this approach and encourages
further development. It also reveals significant uncertainties in the
model and the data. Among these are the uncertainties in the determination
of the start of a solar cycle from the sunspot number series (in particular,
when the cycles overlap), leading to the uncertainty in the phase relation
between the model solution and the data. Also, there are significant
uncertainties in the relationship between the sunspot number data and
the physical properties of the solar magnetic field, in the absence of magnetic
field and helicity data, and, of course, in the dynamo model. Our conclusion
is that for robust and accurate predictions of the solar cycles the information
contained in the sunspot number data is insufficient. For further development
we plan to apply the data assimilation method to more complete 2D dynamo models,
which describe the latitudinal distribution of the solar magnetic field, and
use the magnetograph data available for the past 3 cycles.

This work was supported by the Center for Turbulence Research (Stanford)
and the International Space Science Institute (Bern).



\end{document}